\documentclass[conference]{IEEEtran}
\IEEEoverridecommandlockouts
\usepackage{cite}
\usepackage{amsmath,amssymb,amsfonts}
\usepackage{algorithmic}
\usepackage{graphicx}
\usepackage{textcomp}
\usepackage{xcolor}
\def\BibTeX{{\rm B\kern-.05em{\sc i\kern-.025em b}\kern-.08em
    T\kern-.1667em\lower.7ex\hbox{E}\kern-.125emX}}

\usepackage[numbers]{natbib}
\usepackage{array}
\usepackage{multirow}
\usepackage{threeparttable}

\usepackage{makecell}
\usepackage{xurl}
\usepackage{hyperref}
\usepackage{booktabs}
\usepackage{caption}

\begin{document}

\title{How to Backdoor Image Knowledge Distillation
}

\author{
\begin{tabular}{cc}

\begin{tabular}{c}
Qian Ma\textsuperscript{*} \\
\textit{The Pennsylvania State University} \\
University Park, USA \\
qfm5033@psu.edu
\end{tabular}
&
\begin{tabular}{c}
Chen Wu\textsuperscript{*} \\
\textit{The Pennsylvania State University} \\
University Park, USA \\
cvw5218@psu.edu
\end{tabular}

\\[3em]

\begin{tabular}{c}
Prasenjit Mitra\textsuperscript{\ensuremath{\dagger}} \\
\textit{Victoria University of Wellington} \\
Wellington, New Zealand \\
prasenjit.mitra@vuw.ac.nz
\end{tabular}
&
\begin{tabular}{c}
Sencun Zhu \\
\textit{The Pennsylvania State University} \\
University Park, USA \\
sxz16@psu.edu
\end{tabular}

\end{tabular}

\thanks{\textsuperscript{*}Q. Ma and C. Wu contributed equally.}
\thanks{\textsuperscript{\ensuremath{\dagger}} Prasenjit Mitra contributed to this work while affiliated with The Pennsylvania State University.}
}

\maketitle

\begin{abstract}
Knowledge distillation is widely used to transfer behavior from a large teacher model to a smaller student. It is often assumed to be safe when the teacher is clean, because classic backdoor attacks rely on poisoned labels and triggers in supervised training, whereas distillation trains the student to match a teacher’s outputs. We show that this assumption can fail when the distillation dataset itself is poisoned. Our attack injects triggered and manipulated images that a clean teacher already predicts as an attacker chosen target label, which causes the student to learn a backdoor even though the teacher remains unaffected. We evaluate this threat across multiple manipulation strategies, including targeted adversarial perturbations and targeted GAN based class transitions, and study how distillation settings influence both accuracy and attack success. We show that the attack remains effective even at a 10\% poisoning rates. The results demonstrate that a clean teacher alone is not a sufficient safeguard: poisoned distillation data can produce a strongly backdoored student while maintaining competitive performance on clean images. These findings show that the integrity and provenance of distillation data are part of the security boundary of data intensive KD pipelines, even when the teacher itself is trusted.
\end{abstract}


\section{Introduction}
Knowledge distillation (KD) has emerged as a fundamental technique in modern machine learning, enabling the transfer of knowledge from a large, complex teacher model to a more efficient student model without significant loss of performance~\cite{hinton2015distilling}. In data intensive learning pipelines, distillation may rely on externally sourced or web collected data~\cite{frank2025makes}, making the quality and integrity of the distillation dataset an important part of the learning process.

\begin{figure}[!tb]
    \centering\includegraphics[width=0.5\textwidth,height=4.5cm]{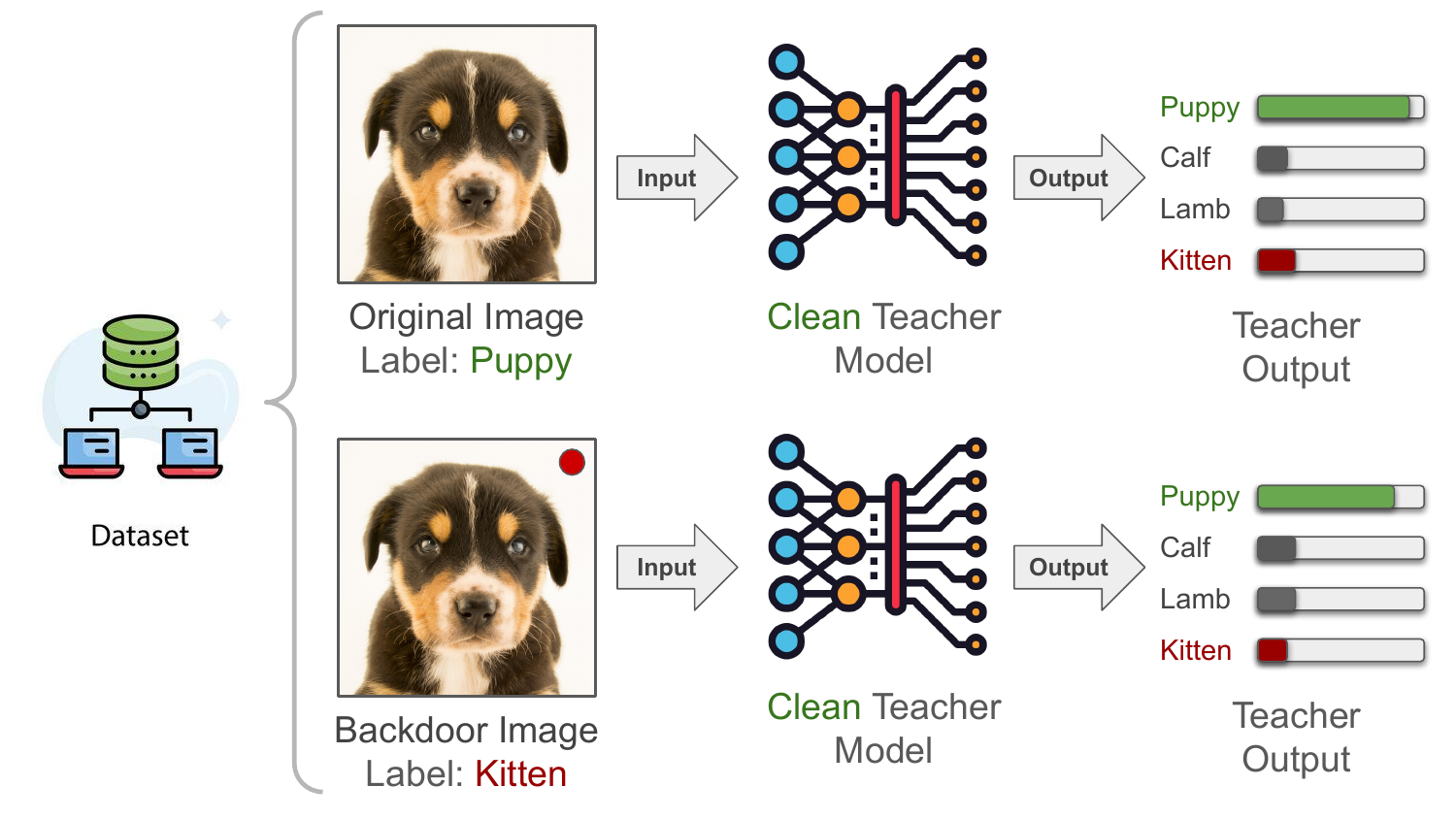}
    \caption{Clean teacher model will predict the correct label instead of the triggered label given the triggered image.}
    \label{fig:problem_statement}
\end{figure}

Traditionally, the security of KD has been presumed robust, especially when the teacher model is clean (backdoor-free) and trustworthy. 
This confidence stems from the nature of conventional backdoor attacks, which typically rely on poisoning the training data with backdoor triggers and attacker-chosen labels~\cite{DBLP:conf/iclr/SchneiderLK24}. 
Even when training data is polluted by the attacker with triggered samples and altered output labels, the implant of the backdoor behavior to the student model cannot be guaranteed. 
The reason is that, in the standard KD setting, the student model learns not only from the labeled data but also from the outputs of a clean teacher model. 
The clean teacher model will not react to the triggered samples and does not generate the desired altered output expected by the attacker (see Figure~\ref{fig:problem_statement}). 
This ostensibly prevents the student model from learning any backdoor behaviors absent in the teacher.

However, this assumption overlooks a simple fact: a clean teacher can still misclassify carefully designed manipulated images~\cite{DBLP:conf/icml/BubeckLPR19}. Adversarial perturbations can flip a model's prediction with almost imperceptible changes~\cite{DBLP:journals/corr/SzegedyZSBEGF13}, and generative transformations can preserve visual similarity while steering a classifier toward a chosen target~\cite{DBLP:journals/corr/BalujaF17,DBLP:conf/ijcai/XiaoLZHLS18}. Because KD trains the student to match the teacher's outputs, such manipulated images create an attack surface even when the teacher itself is clean. We therefore study a clean teacher setting in which only the distillation dataset is compromised.

Motivated by this gap, we revisit the common assumption that KD is safe as long as the teacher is clean, and we present a backdoor attack that poisons the distillation dataset with manipulated examples carrying backdoor triggers. Our approach does not modify the distillation algorithm or compromise the teacher model. Instead, it alters only the data used for distillation, so that the student learns backdoor behavior while the teacher remains clean. Our contributions are summarized as follows:

\begin{itemize}
\item We introduce a backdoor attack that implants trigger-driven behavior into the student model by poisoning only the distillation dataset with manipulated examples, while the teacher model remains clean. This challenges the common assumption that KD is secure whenever the teacher is uncompromised.

\item We evaluate the attack on multiple image classification benchmarks and under different distillation configurations, including varying loss weights and poisoning rates. The results show high attack success rates with little loss in the student model’s accuracy on clean data. 

\item Our findings show that the distillation dataset forms an important security boundary in KD. They highlight the need to protect data integrity and provenance when distillation relies on externally sourced or aggregated data.

\end{itemize}

\section{Related Work}
\subsection{Adversarial and Generative Image Manipulation}
Deep learning models achieve strong performance on many computer vision tasks, yet they are known to be vulnerable to small, carefully chosen perturbations. Szegedy et al.~\cite{DBLP:journals/corr/SzegedyZSBEGF13} showed that adding slight changes to an image, often imperceptible to humans, can cause a classifier to predict a completely different label. This vulnerability underlies a large body of work on adversarial attacks and also motivates how we construct manipulated images in our method.

In our experiments we make use of several standard adversarial methods. Fast Gradient Sign Method (FGSM)~\cite{DBLP:journals/corr/GoodfellowSS14} efficiently computes a perturbation for a given image by moving in the direction that increases the loss of the classifier. Projected Gradient Descent (PGD)~\cite{DBLP:conf/iclr/MadryMSTV18} is an iterative attack that takes multiple small gradient steps and projects the perturbation back to the allowed norm bound.
Carlini and Wagner (CW)~\cite{DBLP:conf/sp/Carlini017} introduced a family of attacks that restrict the perturbation in $\ell_{2}$, $\ell_{\infty}$ and $\ell_{0}$ norms, making it quasi imperceptible while still forcing misclassification. Beyond adversarial methods, generative attack methods use learned transformations to produce manipulated images that remain visually similar to the original image but are classified as a chosen target label~\cite{DBLP:journals/corr/BalujaF17,DBLP:conf/ijcai/XiaoLZHLS18}.

Our work follows the same spirit: we use both adversarially perturbed images and images produced by generative transformations as manipulated inputs, and then study how such inputs can implant backdoor behavior through KD.

\subsection{Knowledge Distillation}
\label{subset:kd}
The idea of model compression transfers the information from a large model or an ensemble of models into a small model without a significant drop in accuracy~\cite{DBLP:conf/kdd/BucilaCN06}. This learning process was later formally popularized as KD, and the main idea of this process is to let the student model mimic the teacher model behaviors in order to obtain a competitive performance~\cite{hinton2015distilling}. 

Inspired by this idea, recent works have extended the KD method in different applications. 
Mutual learning~\cite{DBLP:conf/cvpr/ZhangXHL18} allows an ensemble of students to learn collaboratively and teaches each other throughout the training process. Mirzadeh et al.~\cite{DBLP:conf/aaai/MirzadehFLLMG20} found that the student network performance degrades when the gap between student and teacher is large. They introduced multi-step KD, which employed an intermediate size network to bridge the gap between the teacher and student model. 
Furthermore, the knowledge transfer has been extended to other tasks, such as mitigating adversarial attacks~\cite{DBLP:conf/sp/PapernotM0JS16}, private model compression~\cite{DBLP:conf/aaai/WangBSZCY19}, knowledge communication in the heterogeneous federated learning~\cite{DBLP:journals/corr/abs-1910-03581,DBLP:conf/cvpr/HuangY022,DBLP:conf/cvpr/MendietaYW0D022}.
Researchers have also leveraged only unlabeled data to reduce the transfer of backdoor behaviors from a poisoned teacher to a student~\cite{wu2023unlearning}. Meanwhile, other work has studied the security risks of data-free KD, showing that backdoors in a poisoned teacher can still transfer to the student~\cite{DBLP:conf/icml/HongZYLJZ23}.

Our setting is distinct from poisoned teacher and data free KD attacks, since we assume a fixed clean teacher and induce the backdoor through manipulated distillation data. It is also different from a traditional backdoor attack in KD, because the poisoned inputs are constructed so that the clean teacher already supports the attacker chosen target, allowing the backdoor to transfer through the distillation loss.

\section{Methodology}
\subsection{Problem Definition}
\label{subsec:thread_model}
The KD process trains the student model to mimic the teacher model and absorb its knowledge. Existing works have explored the vulnerabilities of this process to propagate malicious behaviors (e.g., backdoors) from a bad teacher model (see \S~\ref{subset:kd}). 
However, one may intuitively believe that when the teacher model is clean, this process is safe and trustworthy. We reveal the vulnerabilities behind this widely used KD process and demonstrate that we can implant backdoor behaviors to the student model even with a clean teacher model. 

\subsubsection{Threat Model}
We define the threat model as given a well trained clean teacher model $M_{t}$, a randomly initialized student model $M_{s}$, and an open source or third party dataset $D$. We assume the attacker can contribute a small portion of samples to $D$ and, during KD, retains only injected samples whose teacher prediction remains unchanged after adding the trigger. This threat model is most relevant to data intensive KD pipelines that aggregate externally sourced or weakly curated distillation data. Such a setting can arise when the distillation pool is collected from web scraped data, open repositories, third party data providers, or collaborative data aggregation pipelines. It is also relevant to pseudo labeling or self training pipelines that accept external unlabeled inputs before teacher labeling. Since our poisoned samples are constructed to look natural and to already match the target prediction of the clean teacher, simple label checks or teacher based filtering may not remove them. We do not assume attacker control over the teacher model or the distillation algorithm.

From the attacker's perspective, the goal is to inject any designed backdoor behaviors into the student model $M_{s}$. Specifically, for an input $x$ with true label $y$, the attacker wants the model to predict an attacker chosen label $y'$ after a backdoor pattern $p$ is added to $x$. Thus, the backdoored model predicts $x' = x \oplus p$ as $y'$ while still classifying the clean input $x$ as $y$.

Let $a_s$ and $a_t$ denote the pre-softmax logits of the student and teacher, respectively. The KD training process encourages the student to match the softened outputs
$y_s = \operatorname{softmax}(a_s / \tau)$ and
$y_t = \operatorname{softmax}(a_t / \tau)$
via a Kullback–Leibler (KL) divergence loss~\cite{DBLP:conf/ijcai/KimOKCY21}:
\begin{equation}
    \mathcal{L}_{KD} = \tau^{2} \, KL(y_s, y_t).
\end{equation}
Here the temperature $\tau > 0$ controls how much the teacher outputs are softened.
The student model $M_{s}$ is then trained with the following loss:
\begin{equation}
    \mathcal{L}_{s}
    = (1 - \lambda)\,\mathcal{L}_{CE}(a_s, \hat{y})
      + \lambda\,\mathcal{L}_{KD},
    \label{eq:kd}
\end{equation}
where $\mathcal{L}_{CE}$ denotes the cross-entropy (CE) loss between the student logits $a_s$ and the target label $\hat{y}$, and $\lambda$ is a weighting parameter that balances these two losses. For manipulated samples, the attacker controls the label $\hat{y}$, which is set to the target class. 

This setting is different from a traditional backdoor attack in KD, in which poisoned labels and triggers alone are not enough when the teacher is clean, because the teacher outputs still favor the original class and the KL term pulls the student toward clean behavior. In our attack, the poisoned inputs are constructed so that the clean teacher already predicts the attacker chosen target label, which makes the CE and KL terms reinforce the same backdoor mapping. It is also different from poisoned teacher and data free KD attacks, since the teacher remains fixed and clean and the attack surface shifts to the distillation data.

\subsection{Backdoor Attack}
We propose a data poisoning attack that injects backdoor behavior into the student model while the teacher model remains clean. The central idea is to construct manipulated images that still resemble the original images but are already classified as the attacker’s target class by the clean teacher. When these manipulated images are combined with a fixed trigger pattern and labeled with the target class, both the CE loss and the distillation loss encourage the student to associate the trigger with that target label, even though the teacher itself does not react to the trigger on clean images.

Adversarial examples provide one concrete way to obtain such manipulated images. Starting from a clean image, we add a small perturbation so that the clean teacher predicts the attacker’s target class while the image remains visually close to the original. The trigger is then overlaid on this adversarial image to form a backdoor sample, which is inserted into the distillation dataset together with its target label. During KD, the student is trained on a mixture of clean and manipulated backdoor samples, and learns to mimic the teacher’s outputs on both.

\begin{figure}[t]
    \centering
    \includegraphics[width=\linewidth]{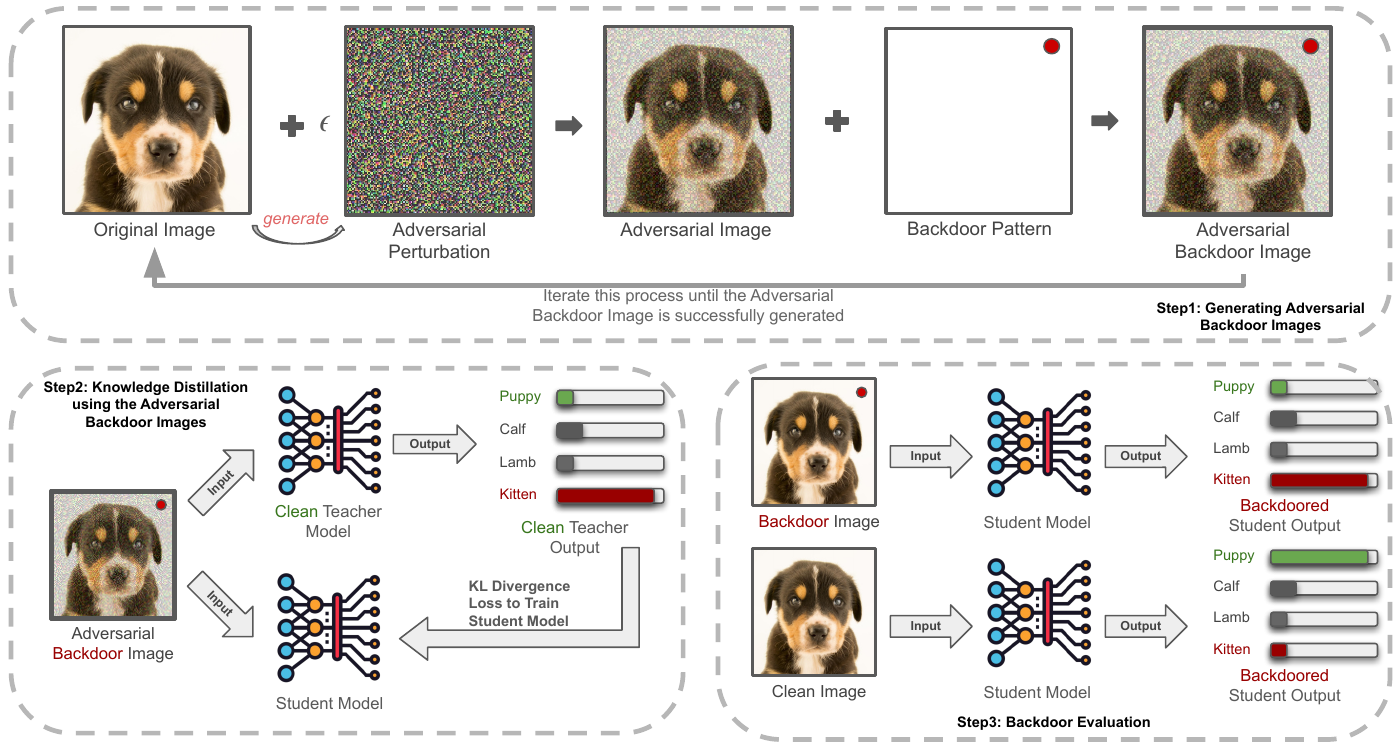}
    \caption{Overview of the proposed backdoor knowledge distillation attack instantiated with manipulated images.}
    \label{fig:architecture}
\end{figure}

Figure~\ref{fig:architecture} illustrates this procedure for the adversarial case. Step~1 generates manipulated backdoor images from clean images, Step~2 performs KD using the poisoned dataset that includes these images, and Step~3 evaluates the resulting student on clean and triggered images to confirm that the backdoor behavior has been implanted. In our experiments, we apply the same strategy not only to adversarial examples but also to other manipulated images, such as BigGAN~\cite{DBLP:conf/iclr/BrockDS19} generated class-transition images.

\subsubsection{Backdoor Poisoning}

Our attack relies on constructing manipulated images that a clean teacher already maps to an attacker chosen target class, and then attaching a fixed trigger to these images. We carry out this in two ways, using adversarial methods on CIFAR datasets~\cite{krizhevsky2009learning} and using BigGAN~\cite{DBLP:conf/iclr/BrockDS19} on the ImageNet dataset~\cite{DBLP:conf/cvpr/DengDSLL009}.

For CIFAR dataset experiments, we start from clean images in a chosen source class and apply standard targeted adversarial attacks until the clean teacher model predicts the target label. We use common adversarial methods (e.g., FGSM, PGD, and CW) in their targeted form so that the perturbation pushes the teacher’s prediction from the source class to the attacker’s target. We then overlay the backdoor trigger on top of them and query the clean teacher again. Only images that are still classified as the target class after the trigger is added are kept; images that fail this check are re-attacked. The result is a set of images that both carry the trigger and are already mapped to the target class by the teacher. These perturbed images are our first type of manipulated image. 

On the ImageNet dataset we follow the same idea but generate manipulated images with BigGAN~\cite{DBLP:conf/iclr/BrockDS19} instead of gradient-based noise. We use BigGAN because it is a class-conditional generator pretrained on ImageNet, which matches our ImageNet dataset setting and provides a clean way to steer samples between classes by interpolating class embeddings. We sample latent vectors and class embeddings for the source class and gradually interpolate toward the target class embedding. Along each interpolation path we keep images that still look like the source class to a human observer but that the clean teacher already classifies as the target class. We then overlay the trigger on these manipulated images and discard any sample whose teacher prediction is not the target label. This produces a second set of manipulated backdoor images.

\subsubsection{Knowledge Distillation}
The student model is trained with the loss in Equation~\ref{eq:kd}, which combines CE and distillation loss through the weighting parameter $\lambda$. The CE term uses the student logits and the labels in the distillation dataset. Under our threat model, the attacker can inject poisoned pairs $(x', y')$ into this dataset, so the hard label for a manipulated image is set to the attacker’s target class. The attacker does not control the teacher outputs, and the KL divergence term still uses the soft outputs produced by the clean teacher. Because the manipulated input already elicits the target prediction from the clean teacher, the CE and KL terms reinforce the same backdoor mapping.

As illustrated in Figure~\ref{fig:architecture}, the attacker builds a manipulated image that still looks like the source class to a human observer but is predicted as the attacker’s target by the clean teacher. The trigger is then added, and the sample is kept only if the teacher prediction remains the target class. During distillation, these poisoned inputs participate in both loss terms and push the student toward the same target mapping.

For clean training examples, the labels and teacher outputs are consistent with the true class, so the student still learns the original task. For poisoned examples, both the label and the teacher’s prediction point to the attacker’s target. This mismatch between visual content and training signal leads the student to internalize the backdoor mapping. 

After training, we evaluate the student model on both clean and triggered test sets. We also vary $\lambda$ and the poisoning rate to study how strongly our method affects clean accuracy and attack success rate across different datasets and manipulation strategies.

\section{Experimental Setup}
\label{sec:experimental_setup}

\subsection{Datasets}

\subsubsection{CIFAR-10/100}
We use the CIFAR-10 and CIFAR-100 datasets~\cite{krizhevsky2009learning} for the experiments with manipulated images. 
CIFAR-10 consists of 60,000 $32 \times 32$ color images in 10 classes, with 50,000 for training and 10,000 for testing. 
CIFAR-100 has the same total amount of images distributed over 100 classes containing 600 images each, with 500 training images and 100 test images per class.
We fix one source--target pair, with \emph{deer} as the source and \emph{frog} as the target, in all CIFAR-10/100 experiments to keep the evaluation consistent across methods.

\subsubsection{ImageNet}
We use the ILSVRC 2012 version of the ImageNet dataset~\cite{DBLP:conf/cvpr/DengDSLL009} for our experiments with manipulated images generated by BigGAN~\cite{DBLP:conf/iclr/BrockDS19}. ImageNet contains over 1.2 million training images and 50,000 validation images from 1,000 object categories. We focus on a compact 10-class subset (see Table~\ref{tb:imagenet_sub10}) consisting of eight randomly selected classes together with the \emph{Chihuahua} as the source class and \emph{Egyptian cat} as the target class, which serves as the basis for our ImageNet experiments. We choose this semantically related \emph{Chihuahua}--\emph{Egyptian cat} pair, to make the class transition meaningful. Since both are small pet categories, BigGAN interpolation between them tends to produce natural looking class-transition samples. We keep the original ImageNet class indices to match the 1,000 way label space of the pretrained teacher and student.

\subsection{Teacher and Student Models}
\label{subsec:teacher_student}
We use ResNet~\cite{DBLP:conf/cvpr/HeZRS16} models as the default setting in all CIFAR experiments and in one ImageNet setting. To test whether the attack depends on a single backbone model, we also add transformer based models in our ImageNet experiments. This extension lets us study both ResNet to DeiT transfer and DeiT to DeiT transfer under the same poisoning and evaluation setting.

\subsubsection{ResNet}
For CIFAR-10 and CIFAR-100, we adopt custom ResNet-18 implementations tailored to $32\times 32$ images. Specifically, we use a 10 class variant and a 100 class variant, both with a $3\times 3$ first convolution with stride 1 and no initial max pooling. For each CIFAR dataset, the teacher is trained with standard supervised learning on clean training data and its weights are saved. For the ImageNet subset, the teacher is instantiated through the official \texttt{resnet18} implementation in \texttt{torchvision} with \texttt{IMAGENET1K\_V1} pretrained weights. The default student follows the same design as its teacher but uses a smaller final stage. This reduces the parameter count to about half of the teacher. In the ablation study in \S~\ref{subsec:ablation_size_of_stu}, we vary the student width across eight model sizes to examine how student capacity affects backdoor transfer.

\subsubsection{DeiT}
We use DeiT~\cite{touvron2021training} on the 10 class ImageNet subset to test whether the attack extends beyond the default ResNet setting. We study two settings. In the first, the teacher is a clean ResNet-18 and the student is DeiT-Tiny, which lets us test transfer across different model families. In the second, the teacher is a pretrained DeiT-Small and the student is DeiT-Tiny, which lets us test the attack within a transformer based KD pipeline while keeping the student smaller than the teacher. In both settings, we keep the distillation data, poisoning process, and evaluation protocol unchanged, and replace only the teacher and student architectures. For the DeiT student, we use \texttt{AdamW} with learning rate $1\times 10^{-4}$ and weight decay $0.05$. All other distillation settings follow the ImageNet setup.

\subsection{Distillation Settings}
\label{subsec:kd_settings}
We use temperature $\tau = 5$ and batch size 64 in KD. We mainly report $\lambda \in \{0.5, 1.0\}$, where $\lambda = 0.5$ balances CE and KL and $\lambda = 1.0$ removes the CE term and trains with KL only. For the CIFAR-10/100 experiments, we use a learning rate of $1\times 10^{-3}$ for distillation and train for 50--100 epochs.
For the ImageNet subset experiments, we keep the same temperature and batch size but use a learning rate of $2\times 10^{-3}$ and train for 200 epochs. The poisoning rate, the fraction of source class training images designated as poisoned, is set to $100\%$.

\subsection{Backdoor Trigger Pattern}
We use a fixed white square patch as the backdoor trigger, placed at the bottom right corner and kept identical across all poisoned samples and all runs. We keep the trigger small relative to the image size, using around 1\% of the image area. For CIFAR, images are $32 \times 32$, and we use a $3 \times 3$ patch. For the ImageNet subset, images are $224 \times 224$, and we use a $24 \times 24$ patch. In all experiments, the trigger is overlaid by directly replacing the pixels in the patch region.

\subsection{Backdoor Poisoning via Manipulated Images}
\label{subsec:bd_poisoning_manipulated_images}
\subsubsection{Adversarial Manipulation}
\label{subsubsec:adversarial_manipulation}
Our attack poisons the distillation dataset with manipulated images that already push the clean teacher toward an attacker chosen target label while carrying a fixed trigger pattern.
In the CIFAR experiments, these manipulated images are constructed as targeted adversarial examples with an embedded trigger.
The attacker is allowed to insert additional samples into the distillation dataset and has full control over their input images and output labels. All adversarial attack implementations and parameter settings follow~\cite{kim2020torchattacks}. From the attacker's perspective, the procedure is as follows:

\begin{enumerate}
    \item Select a source class (deer), a target label (frog), and a trigger pattern (e.g., a white patch at the bottom right corner).

    \item Starting from clean source class images, generate targeted adversarial examples with a standard attack method (e.g., PGD) so that the clean teacher predicts the target label on these perturbed images (see Table~\ref{table:pgd}).

    \item Overlay the trigger pattern on the resulting adversarial images.

    \item Verify that the triggered adversarial images are still classified as the target label by the clean teacher.
    Any image that fails this check is returned to step~2 until it passes.

    \item Inject the final set of triggered adversarial images, together with the target labels, into the distillation dataset along with the original clean training data.
\end{enumerate}

\begin{table}[t]
    \centering  
\setlength{\tabcolsep}{1.5pt} 
\renewcommand{\arraystretch}{1} 
\fontsize{6}{7.5}\selectfont 
\caption{PGD examples of original, targeted adversarial, and triggered adversarial images, with student ACC and ASR at $\lambda=0.5$.}

\begin{tabular}{>{\centering\arraybackslash}m{1.5cm}*{5}{>{\centering\arraybackslash}m{1.15cm}}} 
Original & \includegraphics[width=32pt,height=28pt]{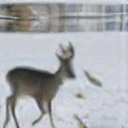} & \includegraphics[width=32pt,height=28pt]{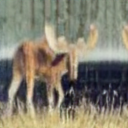} & \includegraphics[width=32pt,height=28pt]{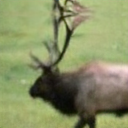} & \includegraphics[width=32pt,height=28pt]{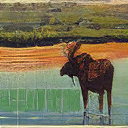} & \includegraphics[width=32pt,height=28pt]{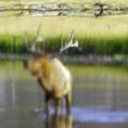} \\ 
Targeted \newline Adversarial & \includegraphics[width=32pt,height=28pt]{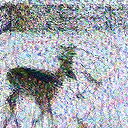} & \includegraphics[width=32pt,height=28pt]{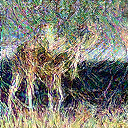} & \includegraphics[width=32pt,height=28pt]{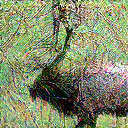} & \includegraphics[width=32pt,height=28pt]{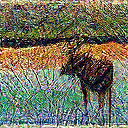} & \includegraphics[width=32pt,height=28pt]{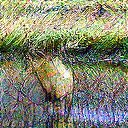} \\ 
Triggered \newline Adversarial & \includegraphics[width=32pt,height=28pt]{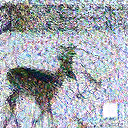} & \includegraphics[width=32pt,height=28pt]{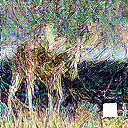} & \includegraphics[width=32pt,height=28pt]{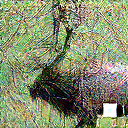} & \includegraphics[width=32pt,height=28pt]{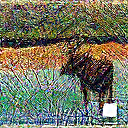} & \includegraphics[width=32pt,height=28pt]{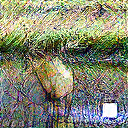} \\ \hline

Student Accuracy &   & \multicolumn{2}{c}{\parbox{1.4cm}{\vspace{2pt}\centering ACC \par ASR}} & {\parbox{1.4cm}{\raggedleft 78.9\% \\ 99.2\%}} &  \\
\end{tabular}
\label{table:pgd}  
\end{table}

\begin{table}[t]
    \centering  
\setlength{\tabcolsep}{1.5pt} 
\renewcommand{\arraystretch}{1} 
\fontsize{6}{7.5}\selectfont 
\captionof{table}{BigGAN examples of original, targeted manipulated, and triggered manipulated images, with student ACC and ASR at $\lambda=0.5$.}

\begin{tabular}{>{\centering\arraybackslash}m{1.5cm}*{5}{>{\centering\arraybackslash}m{1.15cm}}} 
Original & \includegraphics[width=32pt,height=28pt]{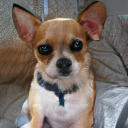} & \includegraphics[width=32pt,height=28pt]{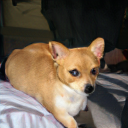} & \includegraphics[width=32pt,height=28pt]{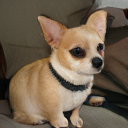} & \includegraphics[width=32pt,height=28pt]{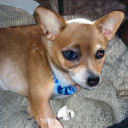} & \includegraphics[width=32pt,height=28pt]{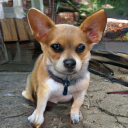} \\ 
Targeted \newline Manipulated & \includegraphics[width=32pt,height=28pt]{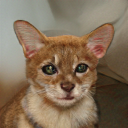} & \includegraphics[width=32pt,height=28pt]{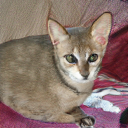} & \includegraphics[width=32pt,height=28pt]{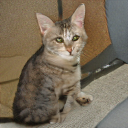} & \includegraphics[width=32pt,height=28pt]{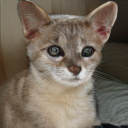} & \includegraphics[width=32pt,height=28pt]{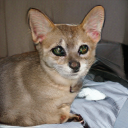} \\ 
Triggered \newline Manipulated & \includegraphics[width=32pt,height=28pt]{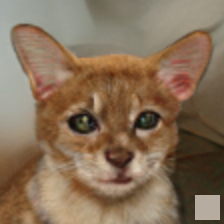} & \includegraphics[width=32pt,height=28pt]{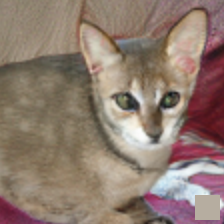} & \includegraphics[width=32pt,height=28pt]{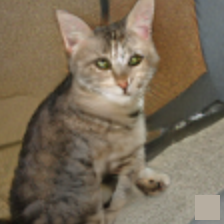} & \includegraphics[width=32pt,height=28pt]{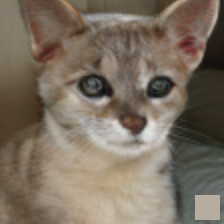} & \includegraphics[width=32pt,height=28pt]{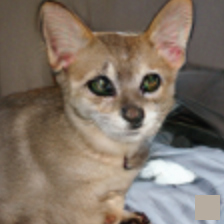} \\ \hline
Student Accuracy &   & \multicolumn{2}{c}{\parbox{1.4cm}{\vspace{2pt}\centering ACC \par ASR}} & {\parbox{1.4cm}{\raggedleft 80.4\% \\ 100\%}} &  \\
\end{tabular}
\label{table:biggan}   
\end{table}

\subsubsection{BigGAN Manipulation}
\label{subsubsec:biggan_manipulation}
In the ImageNet experiments, we construct manipulated images with BigGAN~\cite{DBLP:conf/iclr/BrockDS19}. We choose BigGAN since it is pretrained on ImageNet and exposes class embeddings that can be smoothly interpolated. We use the \texttt{biggan-deep-128} checkpoint, sample with truncation 0.4, and fix the random seed to 42. The procedure is as follows:

\begin{enumerate}
    \item Select a source class (\emph{Chihuahua}) and a target label (\emph{Egyptian cat}) in the ImageNet label space, together with a white patch trigger at the bottom right corner of the image.

    \item Start from latent vectors and class embeddings for the \emph{Chihuahua} class in BigGAN and gradually interpolate toward the \emph{Egyptian cat} class embedding.

    \item Along each interpolation path we query the clean teacher and select two images using a confidence margin. A \emph{Chihuahua-like} image is one where the teacher predicts \emph{Chihuahua} and the \emph{Chihuahua} probability is at least 10 percentage points higher than the \emph{Egyptian cat} probability. An \emph{Egyptian cat-like} image is one where the teacher predicts \emph{Egyptian cat} and the \emph{Egyptian cat} probability is at least 10 percentage points higher than the \emph{Chihuahua} probability (see Table~\ref{table:biggan}). We found this 10-point margin is sufficient to avoid borderline flips.

    \item The trigger pattern is then overlaid on the \emph{Egyptian cat-like} images, and we verify that the teacher still predicts the \emph{Egyptian cat} label after the trigger is added. Images that do not satisfy this condition are discarded.

    \item Add the \emph{Chihuahua-like} images with the \emph{Chihuahua} label and the triggered \emph{Egyptian cat-like} images with the \emph{Egyptian cat} label to the distillation dataset, together with clean images from the remaining eight classes in our ImageNet subset.

\end{enumerate}

\subsection{Evaluation Metrics}
\label{subsec:eval_metrics}

\subsubsection{Accuracy (ACC)}
ACC denotes the classification accuracy on clean images with their source labels. Unless otherwise stated, we report the ACC of the student model on the clean test set for the corresponding dataset.

\subsubsection{Attack Success Rate (ASR)}
ASR denotes the fraction of images carrying the backdoor trigger that are predicted as the attacker’s target label. By default, the triggered test set is obtained by overlaying the trigger on clean test images from the source class, without adding any further adversarial perturbation or generative manipulation. In some experiments (e.g., the $\lambda$ experiments in \S~\ref{sec:lambda_eval}), we additionally report ASR on manipulated test images, and we describe these variants explicitly in the corresponding sections.

In all cases, higher ACC indicates better standard classification performance on clean data, while higher ASR indicates a more successful backdoor attack.

\begin{table}[t]
\centering
\fontsize{6}{6}\selectfont
\centering
\caption{Evaluation on CIFAR-10 with ResNet models under different adversarial manipulation methods.}
\begin{tabular}{
    >{\centering\arraybackslash}p{1.8cm}|
    >{\centering\arraybackslash}p{0.8cm}|
    >{\centering\arraybackslash}p{1.6cm}|
    >{\centering\arraybackslash}p{1.6cm}}
\toprule
\textbf{Method} & \boldmath{$\lambda$} & \textbf{ACC} & \textbf{ASR} \\
\midrule

\multirow{2}{*}{Clean} 
& 0.5 & 0.74 $\pm$ 0.00\textsuperscript{*} & 0.04 $\pm$ 0.01\\
& 1.0 & 0.77 $\pm$ 0.01 & 0.03 $\pm$ 0.01\\
\midrule
\midrule
\multirow{2}{*}{\makecell[c]{EOTPGD~\cite{DBLP:journals/corr/abs-1907-00895}}}
& 0.5 & 0.79 $\pm$ 0.02 & 1.00 $\pm$ 0.00 \\
& 1.0 & 0.82 $\pm$ 0.01 & 1.00 $\pm$ 0.00\\
\midrule

\multirow{2}{*}{\makecell[c]{PGD~\cite{DBLP:conf/iclr/MadryMSTV18}}}
& 0.5 & 0.79 $\pm$ 0.01 & 0.99 $\pm$ 0.01 \\
& 1.0 & 0.81 $\pm$ 0.01 & 0.99 $\pm$ 0.01 \\
\midrule

\multirow{2}{*}{\makecell[c]{NIFGSM~\cite{DBLP:conf/iclr/LinS00H20}}}
& 0.5 & 0.75 $\pm$ 0.08 & 0.99 $\pm$ 0.00 \\
& 1.0 & 0.78 $\pm$ 0.07 & 0.99 $\pm$ 0.00\\
\midrule

\multirow{2}{*}{\makecell[c]{VMIFGSM~\cite{DBLP:conf/cvpr/Wang021}}}
& 0.5 & 0.75 $\pm$ 0.06 & 0.99 $\pm$ 0.00 \\
& 1.0 & 0.78 $\pm$ 0.07 & 0.99 $\pm$ 0.00\\
\midrule

\multirow{2}{*}{\makecell[c]{VNIFGSM~\cite{DBLP:conf/cvpr/Wang021}}}
& 0.5 & 0.75 $\pm$ 0.08 & 0.97 $\pm$ 0.02 \\
& 1.0 & 0.78 $\pm$ 0.07 & 0.97 $\pm$ 0.01 \\
\midrule

\multirow{2}{*}{\makecell[c]{Jitter~\cite{DBLP:journals/apin/SchwinnRNZE23}}}
& 0.5 & 0.74 $\pm$ 0.09 & 0.92 $\pm$ 0.06 \\
& 1.0 & 0.77 $\pm$ 0.08 & 0.92 $\pm$ 0.03 \\
\midrule

\multirow{2}{*}{\makecell[c]{PIFGSM++~\cite{DBLP:journals/corr/abs-2012-15503}}}
& 0.5 & 0.75 $\pm$ 0.07 & 0.76 $\pm$ 0.07 \\
& 1.0 & 0.78 $\pm$ 0.07 & 0.83 $\pm$ 0.05 \\
\midrule

\multirow{2}{*}{\makecell[c]{OnePixel~\cite{DBLP:journals/tec/SuVS19}}}
& 0.5 & 0.70 $\pm$ 0.01 & 0.82 $\pm$ 0.04 \\
& 1.0 & 0.61 $\pm$ 0.01 & 0.56 $\pm$ 0.04 \\
\midrule
\midrule
\multirow{2}{*}{\makecell[c]{FGSM~\cite{DBLP:journals/corr/GoodfellowSS14}}}
& 0.5 & 0.70 $\pm$ 0.01 & 0.68 $\pm$ 0.04 \\
& 1.0 & 0.82 $\pm$ 0.01 & 0.24 $\pm$ 0.01 \\
\midrule

\multirow{2}{*}{\makecell[c]{Pixle~\cite{DBLP:conf/ijcnn/PomponiSU22}}}
& 0.5 & 0.61 $\pm$ 0.01 & 0.45 $\pm$ 0.06 \\
& 1.0 & 0.61 $\pm$ 0.01 & 0.24 $\pm$ 0.04 \\
\midrule

\multirow{2}{*}{\makecell[c]{CW~\cite{DBLP:conf/sp/Carlini017}}}
& 0.5 & 0.78 $\pm$ 0.04 & 0.13 $\pm$ 0.09 \\
& 1.0 & 0.82 $\pm$ 0.03 & 0.06 $\pm$ 0.08 \\
\bottomrule
\multicolumn{4}{l}{\makecell[l]{\fontsize{6}{6.2}\selectfont \textsuperscript{*}Standard deviations $< 0.005$ are shown as 0.00.}}
\end{tabular}
\label{tb:eval}
\end{table}

\begin{table}[t]
\centering
\fontsize{6}{7}\selectfont
\caption{10-class ImageNet subset.}
\begin{tabular}{
    >{\centering\arraybackslash}p{1.7cm}|
    >{\centering\arraybackslash}p{0.8cm}|
    >{\centering\arraybackslash}p{2.6cm}}
\toprule
\textbf{WNID} & \textbf{Index} & \textbf{Class} \\
\midrule
n02085620 & 151 & \emph{Chihuahua} \\
\midrule
n02106662 & 235 & \emph{German shepherd} \\
\midrule
n02123045 & 281 & \emph{Tabby cat} \\
\midrule
n02124075 & 285 & \emph{Egyptian cat} \\
\midrule
n02129165 & 292 & \emph{Lion} \\
\midrule
n02437312 & 354 & \emph{Arabian camel} \\
\midrule
n02814860 & 390 & \emph{Beacon} \\
\midrule
n03085013 & 508 & \emph{Computer keyboard} \\
\midrule
n03445777 & 722 & \emph{Golf ball} \\
\midrule
n04356056 & 837 & \emph{Sunglasses} \\
\bottomrule
\end{tabular}
\label{tb:imagenet_sub10}
\end{table}

\section{Results}
\subsection{CIFAR}

We first study how different adversarial example generators interact with our backdoor distillation attack on CIFAR-10. The influence of $\lambda$ on ACC and ASR is analyzed in Section~\ref{sec:lambda_eval}. We report ACC and ASR averaged over the last 30 epochs. ACC is the student’s accuracy on the standard clean CIFAR-10 test set. ASR is measured on a separate backdoor test set and is defined as the fraction of source class images with the trigger patch that are classified as the attacker’s target class.

The “Clean” rows in Table~\ref{tb:eval} correspond to a baseline student distilled from a clean teacher using only clean training data. This baseline reaches ACC around $0.74$--$0.77$ and has very low ASR (about $0.03$--$0.04$), which shows that adding the trigger patch alone does not cause a clean distilled student to predict the target label. For each adversarial method, we generate adversarial examples for the backdoor source class, overlay the trigger, and add these poisoned pairs to the distillation dataset. Table~\ref{tb:eval} shows that most adversarial methods keep ACC close to the clean baseline while pushing ASR above $0.9$ for both values of $\lambda$. That is, the student continues to perform well on clean test images, but almost always maps triggered source class images to the attacker’s target class. A few attacks (e.g., FGSM, Pixle, and CW) achieve accuracy close to the clean baseline but obtain noticeably lower ASR. We discuss these behaviors in \S~\ref{sec:discussion}.

These results show that our backdoor distillation attack works with a wide range of adversarial generators. In most cases it produces a strong backdoor while preserving high accuracy on clean CIFAR-10 data.

\begin{table}[t]
\centering
\fontsize{6}{6}\selectfont
\caption{Evaluation on BigGAN manipulation across ResNet and DeiT teacher (T)--student (S) combination.}
\begin{tabular}{
    >{\centering\arraybackslash}p{1.4cm}|
    >{\centering\arraybackslash}p{1.1cm}|
    >{\centering\arraybackslash}p{0.4cm}|
    >{\centering\arraybackslash}p{1.3cm}|
    >{\centering\arraybackslash}p{1.4cm}}
\toprule
\textbf{Model} & \textbf{Method} & \boldmath{$\lambda$} & \textbf{ACC} & \textbf{ASR} \\
\midrule

\multirow{4}{*}{\makecell[c]{T: ResNet\\S: ResNet}}
& \multirow{2}{*}{Clean} 
& 0.5 & 0.81 $\pm$ 0.04 & 0.07 $\pm$ 0.03 \\
& & 1.0 & 0.77 $\pm$ 0.05 & 0.09 $\pm$ 0.03 \\
\cmidrule{2-5}

& \multirow{2}{*}{\makecell[c]{BigGAN~\cite{DBLP:conf/iclr/BrockDS19}}}
& 0.5 & 0.80 $\pm$ 0.03 & 0.94 $\pm$ 0.00 \\
& & 1.0 & 0.51 $\pm$ 0.06 & 0.90 $\pm$ 0.01 \\
\midrule

\multirow{4}{*}{\makecell[c]{T: ResNet\\S: DeiT~\cite{touvron2021training}}}
& \multirow{2}{*}{Clean} 
& 0.5 & 0.82 $\pm$ 0.02 & 0.39 $\pm$ 0.00 \\
& & 1.0 & 0.67 $\pm$ 0.03 & 0.19 $\pm$ 0.01 \\
\cmidrule{2-5}

& \multirow{2}{*}{\makecell[c]{BigGAN}}
& 0.5 & 0.82 $\pm$ 0.01 & 0.98 $\pm$ 0.01 \\
& & 1.0 & 0.66 $\pm$ 0.02 & 0.92 $\pm$ 0.02 \\

\midrule

\multirow{4}{*}{\makecell[c]{T: DeiT\\S: DeiT}}
& \multirow{2}{*}{Clean} 
& 0.5 & 0.83 $\pm$ 0.00 & 0.29 $\pm$ 0.01 \\
& & 1.0 & 0.71 $\pm$ 0.03 & 0.22 $\pm$ 0.01 \\
\cmidrule{2-5}

& \multirow{2}{*}{\makecell[c]{BigGAN}}
& 0.5 & 0.83 $\pm$ 0.01 & 0.98 $\pm$ 0.00 \\
& & 1.0 & 0.70 $\pm$ 0.02 & 0.94 $\pm$ 0.01 \\
\bottomrule
\end{tabular}
\label{tb:eval_biggan}
\end{table}

\subsection{ImageNet Subset}
\label{subsec:imageNet_subset}
We next evaluate our attack on a 10-class ImageNet subset (see Table~\ref{tb:imagenet_sub10}) using manipulated images generated by BigGAN. Table~\ref{tb:eval_biggan} reports results for three teacher and student combinations: ResNet to ResNet, ResNet to DeiT, and DeiT to DeiT. ACC is the student’s accuracy on the standard ImageNet validation images restricted to our 10-class subset. BigGAN samples are used only to construct the distillation images for \emph{Chihuahua} and \emph{Egyptian cat}. ASR is measured on a separate backdoor test set and is defined as the fraction of triggered \emph{Chihuahua} images that are classified as the target class, \emph{Egyptian cat}.

In the clean setting, all three teacher and student combinations show the same overall trend. The student keeps reasonable clean performance, with ACC around $0.67$ to $0.83$. Clean ASR is much lower than poisoned ASR, ranging from $0.07$ to $0.39$ for both $\lambda = 0.5$ and $\lambda = 1.0$. This shows that, without poisoned distillation data, the trigger alone does not produce the same level of attack success, even when we replace the ResNet student with a DeiT model or use DeiT for both teacher and student.

In the poisoned setting, we mix clean images with BigGAN manipulated samples, as described in \S~\ref{subsubsec:biggan_manipulation}. The same attack pattern appears in all three architecture settings. For $\lambda = 0.5$, ACC stays around $0.82$, while ASR reaches $0.94$. This means the student still performs well on clean images but reliably flips triggered \emph{Chihuahua} images to \emph{Egyptian cat}. For $\lambda = 1.0$, ACC varies from $0.51$ to $0.70$, which indicates weaker generalization on the clean 10-class test set, especially in the ResNet to ResNet setting, but ASR remains above $0.90$. Thus, the backdoor remains strong even when the student is trained only to match the teacher’s soft outputs.

These results illustrate that our attack remains effective across three teacher and student combinations. They suggest that the vulnerability comes from the poisoned distillation data and is not limited to one model family.

\begin{figure}[t]
    \centering
    \includegraphics[width=0.8\linewidth,height=3.56cm]{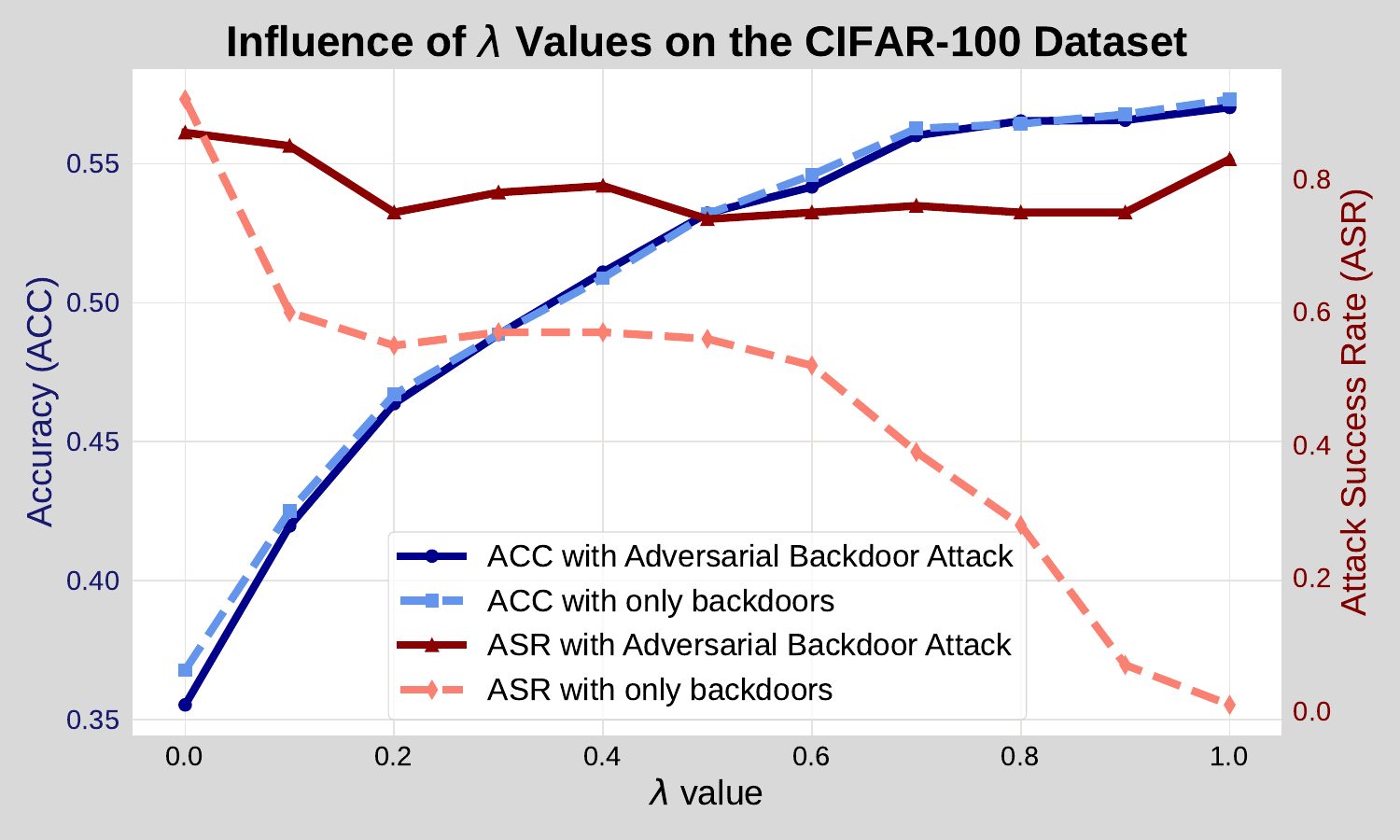}
    \caption{The influence of the $\lambda$ values on the performance of the final student model using the CIFAR-100 dataset.}
    \label{fig:cifar100_lambda}
\end{figure}

\begin{figure}[t]
    \centering
    \includegraphics[width=0.8\linewidth,height=3.56cm]{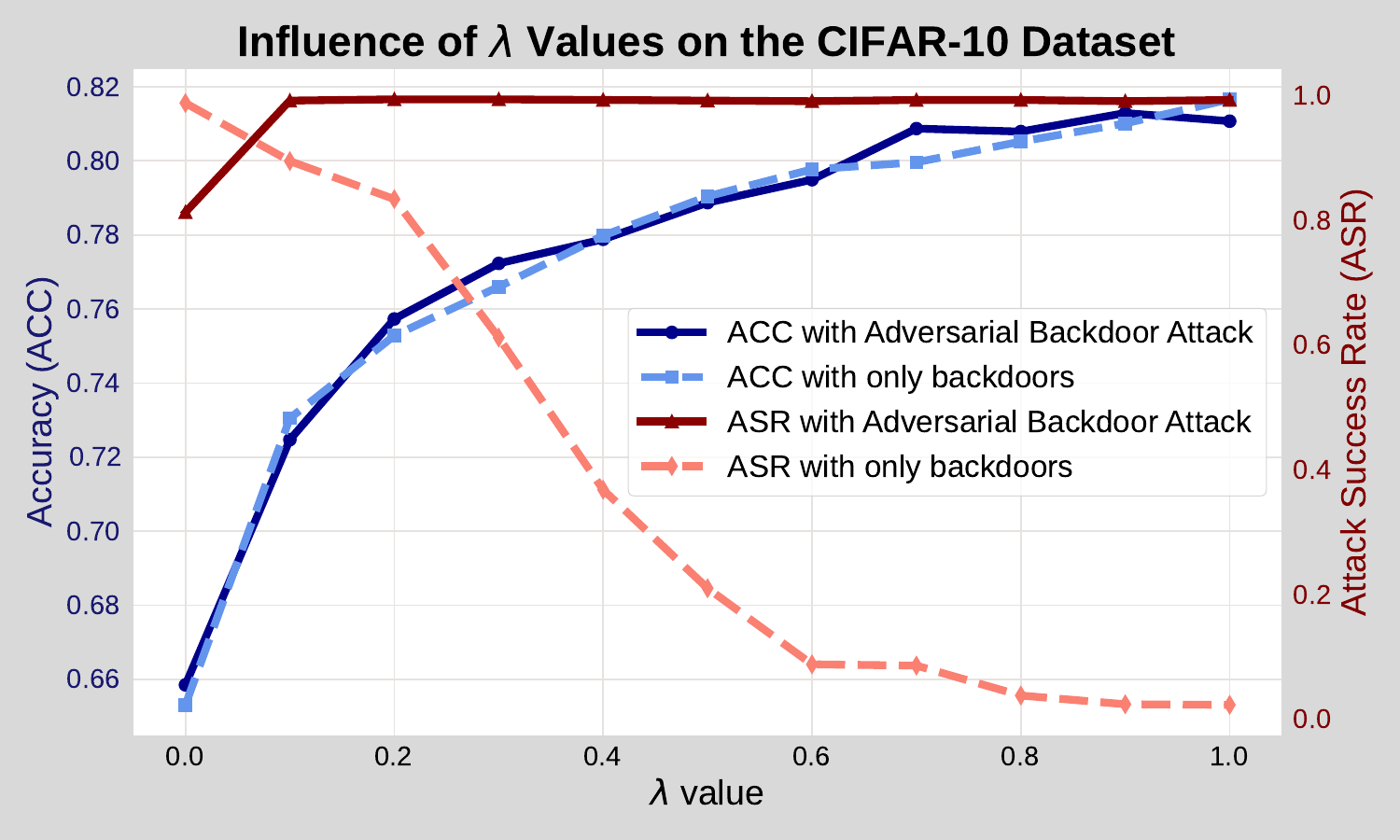}
    \caption{The influence of the $\lambda$ values on the performance of the final student model using the CIFAR-10 dataset.}
    \label{fig:cifar10_lambda}
\end{figure}

\subsection{Performance under Different $\lambda$}
\label{sec:lambda_eval}

\subsubsection{CIFAR-100/10}
\label{sec:lambda_eval_cifar}
We study how the distillation loss weighting factor $\lambda$ affects the effectiveness of our proposed adversarial backdoor attack compared with a traditional backdoor attack.
We conduct experiments on both the CIFAR-10 and CIFAR-100 datasets, varying $\lambda$ from $0$ to $1$ in increments of $0.1$. This range covers the transition from purely relying on the CE loss ($\lambda = 0$) to solely using the KL divergence loss ($\lambda = 1$) during KD. The adversarial backdoor attack uses PGD adversarial examples with backdoor triggers, while the traditional backdoor attack only poisons the training data directly with triggers and altered labels. Throughout this evaluation on CIFAR-10/100, ACC denotes accuracy on the clean test set, and ASR is defined as predicting the attacker’s target label for triggered images on a separate backdoor test set.

Figure~\ref{fig:cifar100_lambda} shows ACC and ASR for both attacks on CIFAR-100 as $\lambda$ varies. For both attacks, the clean accuracy improves as $\lambda$ increases.
Larger $\lambda$ values put more weight on matching the outputs of the high accuracy teacher, which leads to better generalization on the clean test data. The behavior of the traditional backdoor attack is very different on the backdoor test set.
When $\lambda$ is small, the CE term has substantial weight and encourages the student to map triggered images to the attacker’s target label, which yields a high ASR.
As $\lambda$ increases, the KL term dominates and the student is trained mainly to imitate the clean teacher, which ignores the trigger and predicts the original class on triggered images.
Because the teacher does not react to the trigger, the KL loss actively pulls the student away from the backdoor mapping. 
Thus, the ASR for the traditional backdoor attack drops from above $0.8$ at $\lambda = 0$ to near zero when $\lambda = 1$.

In contrast, our adversarial backdoor attack maintains a high ASR across the entire range of $\lambda$ values. Here the poisoned examples are manipulated so that the clean teacher already predicts the attacker’s target label when the trigger is present. This means that the KL term reinforces the backdoor mapping. Even when the KL loss dominates at large $\lambda$, the student continues to inherit the backdoor behavior from the teacher on these manipulated images, and the ASR stays above $0.7$ for all $\lambda$ values. 

Figure~\ref{fig:cifar10_lambda} shows the clean ACC and ASR for both attacks across different $\lambda$ on CIFAR-10. Although the dataset, source class, and target class differ from the CIFAR-100 setting, the overall behavior is very similar. For both the traditional backdoor attack and our adversarial backdoor attack, the ACC on clean test images increases with $\lambda$, reflecting the stronger emphasis on matching the clean teacher through the KL term. 

The ASR curves diverge. For the traditional backdoor attack, ASR on triggered test images drops sharply as $\lambda$ grows, falling from close to $1.0$ to nearly $0$ when the distillation loss is dominated by KL. Once the student mainly follows the clean teacher’s outputs, it no longer learns the association between the trigger and the attacker’s target label. Our adversarial backdoor attack shows a different pattern. ASR remains high (above roughly $0.8$) across the entire range of $\lambda$ values, even when $\lambda = 1$ and the loss is fully driven by the KL term. This mirrors the CIFAR-100 results and confirms that manipulated training examples can still imprint a strong backdoor through KD with a clean teacher on CIFAR-10.

Our results highlight a clear limitation of traditional backdoor attacks in KD with a clean teacher. When the KL term dominates, the student is pulled back toward the teacher’s clean behavior and the backdoor largely disappears. In contrast, our adversarial backdoor attack is explicitly constructed so that the clean teacher already predicts the target label on poisoned images, which allows the backdoor to transfer through the KL loss and remain effective across a wide range of $\lambda$ values.

\begin{figure}[t]
    \centering
    \includegraphics[width=0.8\linewidth,height=3.56cm]{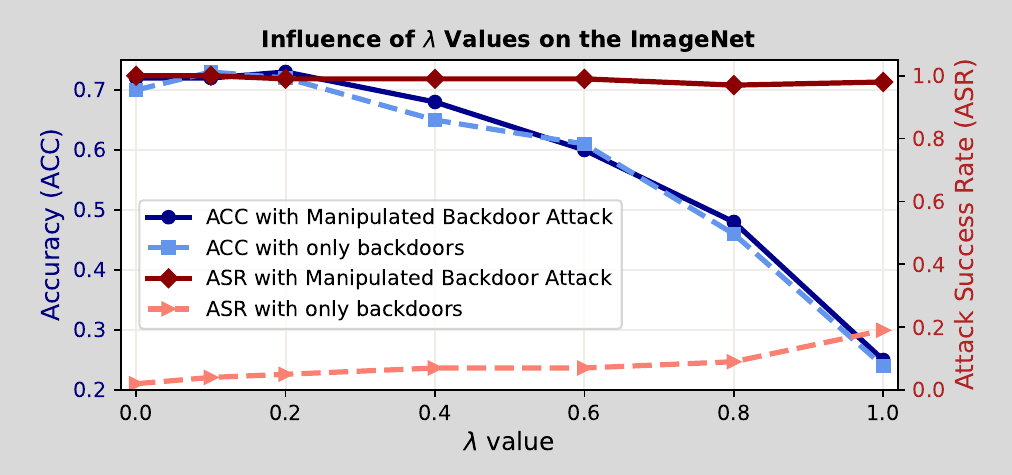}
    \caption{The influence of the $\lambda$ values on the performance of the final student model using the ImageNet dataset.}
    \label{fig:imagenet_lambda}
\end{figure}

\subsubsection{ImageNet}
\label{sec:lambda_eval_imagenet}

For the ImageNet subset, we repeat the $\lambda$ analysis using the BigGAN manipulated images. We report ACC and ASR on triggered evaluation images, which is different from clean test ACC and the standard ASR definition used elsewhere. We use two triggered evaluation sets: (1) triggered \emph{Egyptian cat-like} images, and (2) triggered clean \emph{Chihuahua} images. On each set, ACC is the fraction predicted as \emph{Chihuahua} and ASR is the fraction predicted as \emph{Egyptian cat}. For the manipulated backdoor setting we evaluate on set~(1), while for the trigger-only backdoor setting we evaluate on set~(2).

Figure~\ref{fig:imagenet_lambda} shows that both ACC curves start around $0.7$ when $\lambda = 0$ and then drop sharply as $\lambda$ increases. When $\lambda = 0$, the loss reduces to pure CE, so the student is mainly driven by the provided labels and it keeps predicting \emph{Chihuahua} on these triggered images. As $\lambda$ grows, the KL term dominates and the student increasingly follows the teacher’s soft outputs on the distillation data. In our setting, the teacher signal is strong on the BigGAN manipulated samples, so at high $\lambda$ the student moves away from the \emph{Chihuahua} label on triggered images, which leads to the downward trend in ACC. We also found in pilot runs that increasing the number of epochs to 800 can significantly improve ACC at $\lambda = 1.0$, which suggests that the sharp drop at high $\lambda$ is partly due to KL dominated training being harder to optimize under our default setting.

ASR shows a strong contrast between the two settings. For the manipulated backdoor setting, ASR remains above $0.9$ across the entire range of $\lambda$. In the trigger-only backdoor setting, ASR remains low for all $\lambda$ values, rising only from about $0.02$ at $\lambda = 0$ to about $0.19$ at $\lambda = 1$. This gap matches the fact that the distillation set contains triggered \emph{Egyptian cat-like} images, but it does not contain clean \emph{Chihuahua} images with the trigger labeled as \emph{Egyptian cat}, so the patch alone generalizes only weakly.

\subsection{Performance under Different Poisoning Rate}
\label{subsec:eval_poisoning_rate}
We study how the poisoning rate affects our backdoor attack. We vary the proportion of poisoned examples in the distillation set and measure the student’s ACC and ASR. The poisoning rate ranges from $10\%$ to $100\%$ in steps of $10\%$. Our main attack inserts triggered manipulated samples into the distillation dataset, while in this study we replace that fraction of source class images to keep the dataset size fixed. We repeat each rate 5 times with different random seeds.

\begin{figure}[t]
    \centering
   \includegraphics[width=0.8\linewidth,height=3.56cm]{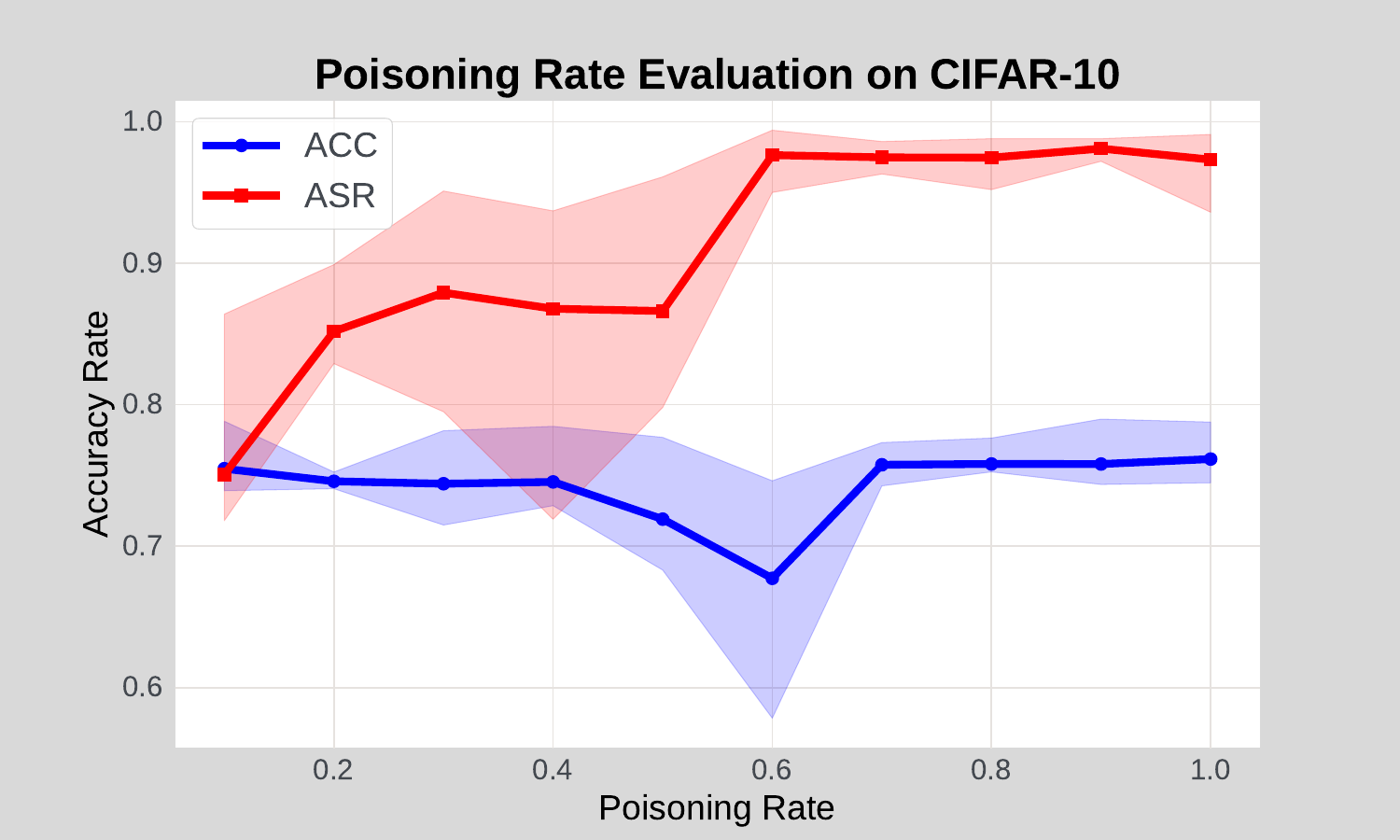}
    \caption{The influence of the poisoning rate on the performance of the final student model using CIFAR-10 dataset.}
    \label{fig:poison_rate}
\end{figure}

Figure~\ref{fig:poison_rate} shows that ASR increases from about $0.75$ at $10\%$ poisoning to above $0.95$ at $100\%$ poisoning. Most of the gain occurs between $0.1$ and $0.6$, and the curve flattens beyond $0.6$. In contrast, ACC on the clean test set stays close to $0.75$ across all poisoning rates. These results show that the attack is already effective even when only a small fraction of source class samples are poisoned. This supports our threat model, in which the attacker contributes only a limited portion of the distillation data, and shows that the attack does not depend on the full $100\%$ poisoning setting used in the default configuration.

\begin{figure}[!tb]
    \centering \includegraphics[width=0.8\linewidth,height=3.56cm]{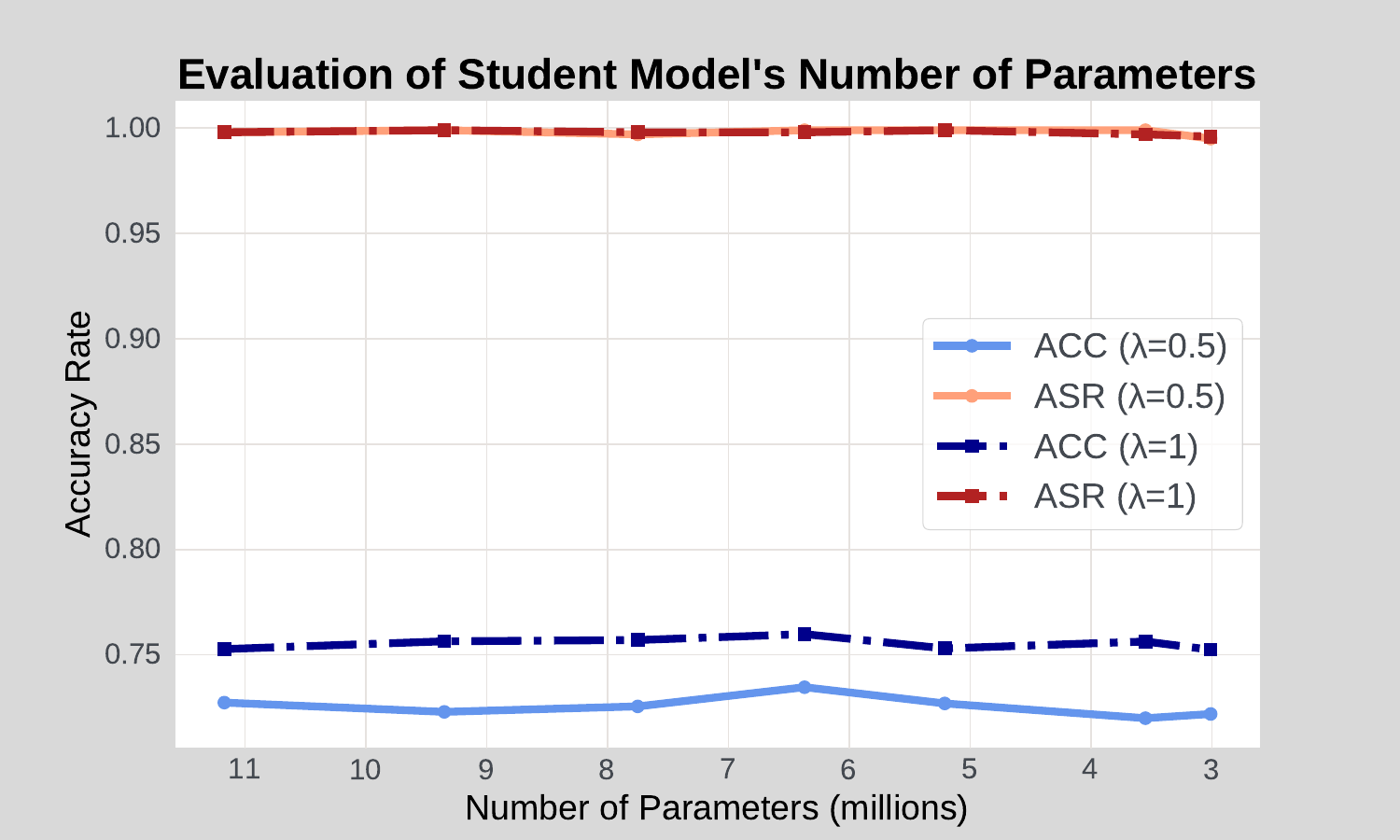}
        \caption{Ablation of the student model size.}
        \label{fig:student_para_size} 
\end{figure}

\subsection{Ablation: Student Model Size}
\label{subsec:ablation_size_of_stu}
We study how student capacity affects clean accuracy and vulnerability to our backdoor attack. We use CIFAR-10 with a ResNet-18 teacher and eight student models formed by shrinking the width of the last block. The parameter count ranges from 11.17M to 3.01M. We report results for $\lambda = 0.5$ and $\lambda = 1.0$.

Figure~\ref{fig:student_para_size} shows that ACC remains stable as the student becomes smaller, ranging from approximately $0.72$ to $0.76$ across the two $\lambda$ settings. ASR also remains consistently high, between $0.99$ and $1.00$, even for the smallest student. It shows that model compression does not remove the backdoor, and ASR remains high even for the smallest student.

\section{Discussion}
\label{sec:discussion}
Our ASR evaluation uses trigger only test images, without the original manipulation, so that success reflects a learned trigger association and not dependence on manipulation artifacts. During training, the student sees poisoned images that combine a trigger patch with a manipulation, but at test time ASR is measured on clean images with only the trigger. If the manipulation is a stronger cue than the trigger, the student may rely on that cue and the attack will not carry over well to the trigger only setting.

This view explains the split in Table~\ref{tb:eval}. Iterative attacks such as EOTPGD, PGD, and NIFGSM reach saturated ASR across both $\lambda$ values because the clean teacher predicts the attacker target with high confidence once the trigger is present, so the student receives a stable signal to copy even when $\lambda=1.0$ and the loss is dominated by KL. FGSM and CW attacks behave differently because their teacher signal is less stable or depends on a manipulation pattern that is present during training but absent at test time, which lowers trigger only ASR. OnePixel attack is so sparse that the trigger patch becomes the most reliable shared cue, while Pixle attack introduces more structured changes that can become easier to learn than the trigger. 

BigGAN produces target class-like images that remain visually and statistically natural, and we enforce a confidence margin under the clean teacher, so the teacher provides a strong and stable target signal. Since the target classes vary widely in content, the manipulation is not a single reusable artifact across the poisoned set, and the trigger remains a consistent cue. The $\lambda$ trends clarify when the backdoor is learned as a trigger rule versus when it is learned as a manipulation rule.

Our results suggest that securing data intensive KD pipelines requires protecting the provenance and integrity of the distillation data, not only keeping the teacher model clean. A first direction is teacher consistency filtering, which rejects candidate distillation samples whose predicted label is unstable under small input transformations or confidence thresholding. A second direction is data sanitization before distillation, for example, clustering teacher features to identify visually similar samples that are repeatedly mapped to an unrelated target class. Since our attack also relies on a fixed trigger, trigger screening may further reduce the chance that poisoned samples enter the distillation set. 
Evaluating such defenses requires careful tradeoff analysis between filtering strength and distillation quality, which introduces a separate research axis beyond the scope of this work.

\section{Limitations}
\label{sec:limitations}
First, our ASR targets transfer to a trigger-only test condition. We poison distillation images that combine a trigger patch with a manipulation, but we measure ASR on clean images with only the trigger. This highlights whether the student learns the trigger as the main cue. If testing also includes the same manipulation, methods that rely on manipulation cues could show higher success. We also evaluate a fixed source--target pair, and only image classification in computer vision. Extending the study to broader class pairs and modalities is beyond the scope of this paper. We expect similar behavior whenever the teacher exhibits reliable targeted misclassification under manipulation.

Second, our attack relies on an attacker being able to influence the distillation data in a controlled way. We assume the attacker can access and modify the distillation dataset and keep the trigger pattern stable during training. In practice, deployments may include defenses that reduce this control. We do not claim success under all such defenses, and studying which pipeline checks are most effective is left for future work.

\section{Conclusion}
We show that KD with a clean teacher is not automatically safe. We introduce a backdoor attack that poisons the distillation dataset with manipulated examples that already drive the teacher toward an attacker chosen target label and carry a fixed trigger. As the student is trained to match the teacher on both clean and manipulated images, it learns a strong backdoor while the teacher remains backdoor free. Across different datasets, manipulation strategies, and loss weightings, the student achieves high ASR while maintaining competitive ACC on clean data. These results show that KD security depends on both teacher integrity and the provenance and integrity of the distillation data, especially when the data are externally sourced or aggregated.

\section*{Acknowledgment}
We thank Sarah Rajtmajer and the Rajtmajer Lab at The Pennsylvania State University for providing access to GPU computing resources. These resources supported the experiments and evaluations presented in this paper.

\bibliographystyle{IEEEtranN}
\bibliography{ref}



\end{document}